\documentclass[10pt,twocolumn]{article}
\usepackage[a4paper,margin=1.7cm]{geometry}
\usepackage{graphicx}
\usepackage{tikz}
\usetikzlibrary{positioning,arrows.meta,shapes,calc}
\usepackage{booktabs}
\usepackage{microtype}
\usepackage{url}
\usepackage{orcidlink}

\definecolor{orangeAccent}{RGB}{255,128,50}
\definecolor{mintAccent}{RGB}{170,255,240}
\definecolor{violetAccent}{RGB}{190,130,255}
\definecolor{blueAccent}{RGB}{90,150,255}
\definecolor{tealAccent}{RGB}{130,255,210}

\title{Toward Automated Virtual Electronic Control Unit (ECU) Twins for Shift-Left Automotive Software Testing}
\author{Sebastian Dingler\orcidlink{0000-0002-0162-8428}\thanks{Sebastian Dingler conducted this work as independent research for the University of Applied Sciences Esslingen, Flandernstr.\ 101, 73732 Esslingen, Germany. For correspondence, remove \texttt{foobar} from \texttt{s.dinglerfoobar@gmail.com}.}\and Frederik Boenke\\NUVUS GmbH}
\date{}

\begin{document}
\pagestyle{myheadings}
\markboth{\tiny THIS WORK WILL BE SUBMITTED TO 39. VDI-Tagung Fahrerassistenzsysteme und Automatisiertes Fahren FOR POSSIBLE PUBLICATION. COPYRIGHT MAY BE TRANSFERRED WITHOUT NOTICE.}{\tiny THIS WORK WILL BE SUBMITTED TO 39. VDI-Tagung Fahrerassistenzsysteme und Automatisiertes Fahren FOR POSSIBLE PUBLICATION. COPYRIGHT MAY BE TRANSFERRED WITHOUT NOTICE.}
\maketitle
\thispagestyle{myheadings}

\begin{abstract}
Automotive software increasingly outpaces hardware availability, forcing late integration and expensive hardware-in-the-loop (HiL) bottlenecks. The InnoRegioChallenge project investigated whether a virtual test and integration environment can reproduce electronic control unit (ECU) behavior early enough to run real software binaries before physical hardware exists. We report a prototype that generates instruction-accurate processor models in SystemC/TLM~2.0 using an agentic, feedback-driven workflow coupled to a reference simulator via the GNU Debugger (GDB). The results indicate that the most critical technical risk---CPU behavioral fidelity---can be reduced through automated differential testing and iterative model correction. We summarize the architecture, the agentic modeling loop, and project outcomes, and we discuss the technical approach in a manner consistent with the reported qualitative findings. While cloud-scale deployment and full toolchain integration remain future work, the prototype demonstrates a viable shift-left path for virtual ECU twins, enabling reproducible tests, non-intrusive tracing, and fault-injection campaigns aligned with safety standards.
\end{abstract}


\section{Introduction}
Hardware scarcity, long lead times, and binary-only deliveries from suppliers constrain modern automotive software development. Virtual electronic control units (vECUs)---also commonly referred to as virtual platforms in this context---are software-based replicas of electronic control units (ECUs) that execute production binaries while emulating the processor, memory map, peripherals, and in-vehicle communication. A typical vECU or virtual platform comprises an instruction-set simulator, models of peripherals, accelerators, and other hardware components, models of the communication infrastructure, and an integration framework that hooks these elements together and advances simulation time, for example SystemC/TLM. In 2025, the InnoRegioChallenge 2025 project, funded by the Wirtschaftsf\"orderung Region Stuttgart GmbH (WRS), explored whether vECUs can reproduce ECU behavior early enough to support shift-left testing before hardware prototypes are available. The primary objective was to validate technical feasibility rather than deliver a production-grade cloud platform.

vECUs enable early software integration, regression testing, and safety analysis when physical hardware is unavailable. However, different abstraction levels trade timing fidelity against performance and observability, and the appropriate point in this design space depends on the intended use (e.g., interactive debugging, continuous integration, or safety validation).

Timing is not merely a performance detail: safety arguments and regulatory requirements for automated driving and advanced driver-assistance systems (ADAS) depend on bounded end-to-end latencies (e.g., event-chain timing) that must be evaluated and verified~\cite{dingler2025eventchain}.

At the implementation level, vECUs often combine two complementary technologies. SystemC, standardized as IEEE 1666, provides a C++-based modeling language with a discrete-event simulation kernel; its Transaction-Level Modeling (TLM) library abstracts communication as timed transactions (e.g., reads, writes, and bursts) rather than signal-accurate toggling, improving simulation speed while retaining ordering and timing control~\cite{systemc}. SystemC can host functional or instruction-accurate CPU models in C++, but instruction decoding and execution may become the dominant cost when running large software stacks.

QEMU, by contrast, is an instruction-set simulator that can run either via interpretation or dynamic binary translation (DBT). In DBT mode, QEMU decodes guest instructions, translates basic blocks at runtime into host code through an intermediate representation, and caches translated blocks for reuse~\cite{qemu}. QEMU's Tiny Code Generator (TCG) implements this just-in-time compilation pipeline by lowering guest instructions into a host-independent intermediate form and emitting host machine code for cached translation blocks~\cite{qemu}. Many timing-aware (approximately timed) vECUs therefore adopt a hybrid architecture: QEMU provides fast instruction-set architecture (ISA) execution, while SystemC/TLM composes the surrounding timed platform model; timing fidelity is then enforced by explicit synchronization between QEMU execution and SystemC time (e.g., instruction counting or periodic time annotation)~\cite{dvcon2023qemu}. SystemC emphasizes compositional modeling and timing control, whereas QEMU emphasizes execution speed and broad ISA support.

The contributions of this paper are fourfold:
\begin{itemize}
  \item[(i)] a structured approach from ECU teardown and artifact analysis to a virtual twin, mapping digital, standard-IC, and analog elements to suitable modeling formalisms;
  \item[(ii)] an agentic two-loop calibration workflow in which a generative coding model synthesizes and iteratively repairs SystemC/TLM simulation code using deterministic feedback;
  \item[(iii)] an instruction-level, GDB-based differential validation strategy that compares generated CPU behavior against a trusted reference simulator at the architecturally visible ISA level; and
  \item[(iv)] a prototype case study that demonstrates the feasibility of automated CPU-model generation for virtual ECU twins and distills practical lessons for shift-left automotive software testing.
\end{itemize}

\section{Background}
Seen from a classical top-down design flow, simulation spans a hierarchy of abstractions. At the system level, requirements and functions are captured in natural language, UML, C++, or similar high-level notations. The next refinement step is the algorithmic or transaction level, where concurrent functional blocks communicate through untimed, loosely timed, or approximately timed transactions. This is the design space in which SystemC/TLM is particularly effective, because it preserves software-visible behavior and architectural structure without forcing signal-accurate implementation details~\cite{systemc,dvcon2023tutorial}.

Refining the model further leads to register-transfer level (RTL) descriptions, where pin- and cycle-accurate behavior, clocked data flow, and bus interfaces become explicit. Below RTL, gate-level models focus on Boolean structure and sequential logic built from library cells, while switch-level and electrical-level models expose transistor and circuit behavior with increasing physical realism. These lower levels are essential for implementation sign-off, signal integrity, or analog effects, but they are substantially more expensive to simulate; mixed-signal extensions such as SystemC-AMS are therefore typically reserved for components whose electrical behavior materially affects system function~\cite{li}. For vECUs, the practical sweet spot is usually above gate level: high enough to execute real software stacks efficiently, but detailed enough to retain software-visible timing, ordering, and interface semantics~\cite{dvcon2023tutorial}.

\section{Related Work}

Related work on virtual platforms and vECUs spans (i) fault injection and fault-effect analysis on SystemC/TLM virtual prototypes, (ii) mixed-signal or AUTOSAR-oriented verification setups, and (iii) performance and observability tooling for running large software stacks.

Early work on SystemC/TLM virtual prototypes established non-intrusive fault injection and fault-effect analysis as enablers for shift-left robustness studies. Tabacaru \emph{et al.} propose fault-injection techniques for TLM-based virtual prototypes~\cite{tabacaru}. Li \emph{et al.} describe a SystemC/SystemC-AMS virtual prototyping, verification, and validation framework for automotive systems, emphasizing functional verification rather than automated peripheral modeling or ISO~26262 (functional safety)-aligned fault campaigns~\cite{li}. Tabacaru's doctoral thesis studies fault-effect analysis at the virtual-prototype abstraction level, supporting shift-left safety analysis while leaving open workflows for building complete ECU twins and refining peripheral behavior at scale~\cite{fae}.

Performance and scalability are prominent themes in more recent work. Bosbach \emph{et al.} parallelize CPU models while keeping the SystemC kernel standard-compliant, achieving multi-core speedups while reinforcing timing-accuracy trade-offs associated with temporal decoupling~\cite{dac2024}. J\"unger \emph{et al.} report a fast RISC-V virtual platform based on a custom just-in-time compilation engine and show that higher performance and instruction-accurate timing are mutually exclusive in standard QEMU-style execution~\cite{dvcon2022}.
As an alternative to dynamic binary translation, Bosbach \emph{et al.} demonstrate ARM-on-ARM virtualization for SystemC/TLM virtual platforms using Linux's Kernel-based Virtual Machine (KVM), executing target software natively on Arm hosts and reporting substantial speedups over instruction-set-simulator-based CPU models~\cite{bosbach2025aoa}.

Tutorial and framework contributions provide context for vECU design choices. Bosbach, J\"unger, and Leupers outline abstraction levels (analog to loosely timed), explicitly linking higher performance to reduced timing fidelity~\cite{dvcon2023tutorial}. J\"unger \emph{et al.} present a framework that combines QEMU with SystemC TLM-2.0 to build full-system vECUs with vehicle-network integration, illustrating synchronization complexity and engineering overhead~\cite{dvcon2023qemu}. Kim \emph{et al.} report an AUTOSAR-compatible Level-4 vECU for cloud-native verification, demonstrating close task-timing alignment to target ECUs and automated testing workflows~\cite{electronics2024}.

Complementary to VP fidelity and performance, interoperability across heterogeneous toolchains is addressed by Pollo \emph{et al.}, who automatically wrap SystemC models into Functional Mock-up Units (FMUs) following the Functional Mock-up Interface (FMI) standard to enable portable co-simulation and model encapsulation for software-defined vehicle workflows~\cite{pollo2025fmi}.
Along similar lines, Bosbach \emph{et al.} present an FMI-based framework that exposes a SystemC/TLM virtual platform through an adapter FMU, enabling cross-tool co-simulation (e.g., environmental models and test tools) while keeping the virtual platform and target software unmodified~\cite{bosbach2025fmi_systemc}.

Finally, domain-specific work underscores the challenge of test throughput for safety workloads. Profiling of Level~4 vECUs for faster ISO~26262 testing reports large speedups after bottleneck identification, indicating substantial optimization effort is often required~\cite{dvcon2024iso}. Bosbach \emph{et al.} introduce non-intrusive QEMU code coverage for embedded systems, demonstrating how host-side tooling enables deeper verification while adding performance overhead~\cite{rapido2024nqc2}. J\"unger \emph{et al.} show that CPU model efficiency is a primary driver of system-level simulation speed through Unicorn-based SystemC processors~\cite{rapido2019unicorn}.

In contrast to these works, which predominantly rely on manually engineered models, vendor-provided virtual platforms, or conventional framework integration, this paper investigates whether a feedback-driven LLM agent can synthesize and self-correct ISA-level CPU models against a trusted reference simulator. The contribution is therefore not another manually constructed vECU model, but a generative calibration workflow for producing and iteratively repairing architecturally consistent simulation behavior.

\section{Approach: From ECU Teardown to Virtual Twin}
The project defined a systematic path from physical ECU artifacts to a virtual twin, starting from teardown-based analysis of layout, schematics, and bill of materials. Identified components are mapped to appropriate modeling formalisms:
\begin{itemize}
  \item \textbf{Digital compute and buses} modeled in SystemC/TLM~2.0 (instruction execution, memory map, bus transactions).
  \item \textbf{Standard ICs} modeled from datasheets (register maps, timing constraints).
  \item \textbf{Analog components} modeled via SPICE/SystemC-AMS equivalents (filters, power integrity effects).
\end{itemize}

In this paper, the emphasis is on the digital-compute path (CPU modeling).

\section{Method: Agentic Two-Loop Calibration}
The presented workflow follows a ``generate--evaluate--revise'' pattern for simulation code. It is inspired by NVIDIA Research's work ``EUREKA: Human-Level Reward Design via Coding Large Language Models,'' which iteratively improves LLM-generated program artifacts via structured feedback derived from training metrics~\cite{eureka}. Whereas EUREKA uses reinforcement-learning training metrics as feedback, our setting optimizes deterministic SystemC/TLM model code using model-fidelity signals such as trace deltas, timing deviations, and fault responses. In the implementation reported here, Claude Opus~4-class coding models were used as constrained synthesis engines for SystemC/TLM code rather than as sources of semantic ground truth. Correctness remained anchored in ISA semantics, reference execution, and deterministic comparison outcomes.

\paragraph{Loop A -- Generative Model Synthesis.} Based on available system artifacts---interface specifications, register and memory maps, communication descriptions, reference traces, and optional architecture or AUTomotive Open System ARchitecture (AUTOSAR) information---a generative model produces initial hypotheses of structural and behavioral components. These include SystemC/TLM peripheral models, SoC glue logic, and parameterized assumptions about timing, register logic, and communication behavior. In the CPU case study, the initial task formulation instructed the model to build a simulator together with an automated comparison harness against a trusted reference model. The workflow was semi-supervised: after this initial task definition, the model generated most of the implementation autonomously, while a human operator intervened only when the search stalled, entered a non-convergent implementation path, or required decomposition into smaller repair tasks.

\paragraph{Loop B -- Deterministic Feedback Loop.} The generated model is executed deterministically and compared against reference behavior. Metrics include register- and bus-level trace deltas, timing deviations, state-transition mismatches, resource profiles, and fault responses under injected disturbances. These measures are aggregated into a multi-dimensional score that guides the next iteration of model refinement. At the instruction level, the feedback exposed concrete divergences such as unexpected register contents, incorrect condition flags, erroneous program-counter progression, or instruction-visible memory effects. Candidate revisions were retained only if they improved agreement with the reference behavior under deterministic replay.

The resulting implementation was not obtained in a single generation step. Instead, the model repeatedly proposed candidate code, observed discrepancy reports derived from comparison against the reference simulator, and revised the affected logic until the generated behavior converged toward the specified architectural semantics. In this sense, the generative model primarily accelerated hypothesis formation and local repair, while the reference-guided evaluation loop enforced correctness.

\begin{figure}[t]
\centering
\resizebox{\columnwidth}{!}{%
\begin{tikzpicture}[font=\scriptsize, node distance=7mm, x=1cm, y=1cm,
  box/.style={draw, rounded corners=5pt, line width=0.7pt, align=center, inner sep=4pt},
  accent/.style={-Latex, line width=0.8pt},
]
  \node[box, fill=orangeAccent!12, draw=orangeAccent!70] (artifacts) {System\\artifacts};
  \node[box, right=1.2cm of artifacts, minimum width=3.2cm, fill=violetAccent!10, draw=violetAccent!70] (loopa) {Loop A\\Generative model synthesis};
  \node[box, right=1.2cm of loopa, minimum width=2.6cm, fill=black!5, draw=black!50] (model) {Candidate\\model};
  \node[box, below=0.9cm of model, minimum width=3.2cm, fill=blueAccent!12, draw=blueAccent!70] (loopb) {Loop B\\Feedback evaluation};
  \node[box, below=0.9cm of loopa, minimum width=2.6cm, fill=mintAccent!12, draw=mintAccent!70] (metrics) {Deterministic\\metrics};

  \draw[accent, color=orangeAccent!80] (artifacts) -- (loopa);
  \draw[accent, color=violetAccent!80] (loopa) -- (model);
  \draw[accent, color=blueAccent!80] (model) -- (loopb);
  \draw[accent, color=blueAccent!80] (loopb) -- (metrics);
  \draw[accent, color=tealAccent!80] (metrics.north) -- (loopa.south);
\end{tikzpicture}
}
\caption{Two-loop calibration cycle: Loop~A synthesizes candidate models from available artifacts; Loop~B evaluates them using deterministic metrics and feeds the results back for refinement.}
\label{fig:loops}
\end{figure}
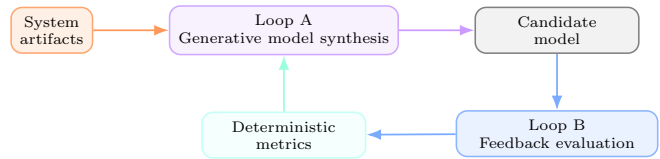

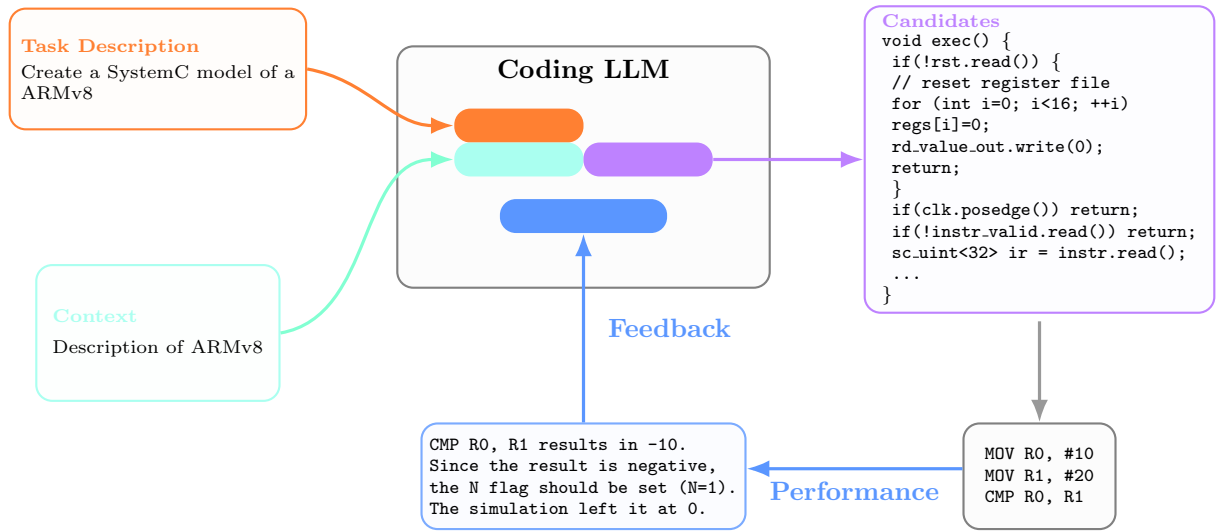
\begin{figure*}[t]
\centering
\begin{tikzpicture}[font=\small, node distance=8mm, x=1cm, y=1cm,
  box/.style={draw, rounded corners=6pt, line width=0.8pt, align=left, inner sep=6pt, minimum height=1.1cm},
  darkbox/.style={box, fill=white, text=black},
  bubble/.style={rounded corners=6pt, minimum height=0.45cm},
  accent/.style={line width=1.2pt, -{Latex[length=3mm]}},
]


\node[darkbox, minimum width=3.9cm, minimum height=1.6cm, draw=orangeAccent, fill=orangeAccent!10, fill opacity=0.15, text opacity=1, font=\scriptsize, inner sep=4pt] (task) at (2.6,6.7) {
  {\bfseries\color{orangeAccent}Task Description}\\[2pt]
  Create a SystemC model of a\\
  ARMv8
};

\node[darkbox, minimum width=3.2cm, minimum height=1.8cm, draw=mintAccent, fill=mintAccent!10, fill opacity=0.15, text opacity=1, font=\scriptsize, inner sep=3pt] (context) at (2.6,3.2) {
  {\bfseries\color{mintAccent}Context}\\[4pt]
  Description of ARMv8
};

\node[darkbox, minimum width=4.9cm, minimum height=3.2cm, draw=black!50, fill=black!10, fill opacity=0.12, text opacity=1, align=center] (llm) at (8.2,5.4) {};
\node[font=\bfseries, text=black] (llm_title) at ($(llm.north)+(0,-0.35cm)$) {Coding LLM};

\node[bubble, fill=orangeAccent, minimum width=1.7cm] (bubble_task) at ($(llm.center)+(-0.85cm,0.55cm)$) {};
\node[bubble, fill=mintAccent, minimum width=1.7cm] (bubble_context) at ($(llm.center)+(-0.85cm,0.10cm)$) {};
\node[bubble, fill=violetAccent, minimum width=1.7cm] (bubble_candidates) at ($(llm.center)+(0.85cm,0.10cm)$) {};
\node[bubble, fill=blueAccent, minimum width=2.2cm] (bubble_feedback) at ($(llm.center)+(0.00cm,-0.65cm)$) {};

\node[darkbox, minimum width=4.6cm, minimum height=3.2cm, draw=violetAccent, fill=violetAccent!10, fill opacity=0.15, text opacity=1, font=\scriptsize, inner sep=3pt, align=left] (candidates) at (14.2,5.5) {
  {\bfseries\color{violetAccent}Candidates}\\[-1pt]
  \texttt{void exec() \{}\\
  \texttt{ if(!rst.read()) \{}\\
  \texttt{  // reset register file}\\
  \texttt{  for (int i=0; i<16; ++i)}\\
  \texttt{   regs[i]=0;}\\
  \texttt{  rd\_value\_out.write(0);} \\
  \texttt{  return;}\\
  \texttt{ \}}\\
  \texttt{ if(clk.posedge()) return;}\\
  \texttt{ if(!instr\_valid.read()) return;}\\
  \texttt{ sc\_uint<32> ir = instr.read();}\\
  \texttt{ ...}\\
  \texttt{\}}
};

\node[box, fill=blueAccent!12, fill opacity=0.18, text opacity=1, text=black, minimum width=4.0cm, minimum height=1.4cm, draw=blueAccent!80, font=\scriptsize, inner sep=3pt] (feedback) at (8.2,1.3) {
  \texttt{CMP R0, R1 results in -10.}\\
  \texttt{Since the result is negative,}\\
  \texttt{the N flag should be set (N=1).}\\
  \texttt{The simulation left it at 0.}
};

\node[box, fill=black!10, fill opacity=0.12, text opacity=1, text=black, minimum width=2.0cm, minimum height=1.4cm, draw=black!50, font=\scriptsize, inner sep=2pt] (perf) at (14.2,1.3) {
  \texttt{MOV R0, \#10}\\
  \texttt{MOV R1, \#20}\\
  \texttt{CMP R0, R1}
};

\coordinate (fb_up) at ($(feedback.north)+(0,0.0)$);
\coordinate (fb_right) at ($(feedback.east)+(0,0.1)$);
\coordinate (perf_left) at ($(perf.west)+(0,0.1)$);
\coordinate (cand_down) at ($(candidates.south)+(0,-0.1)$);
\coordinate (perf_up) at ($(perf.north)+(0,0.0)$);

\draw[accent, color=orangeAccent] (task.east) to[out=-10, in=180] (bubble_task.west);
\draw[accent, color=tealAccent] (context.east) to[out=10, in=180] (bubble_context.west);
\draw[accent, color=violetAccent] (bubble_candidates.east) -- (candidates.west);
\draw[accent, color=blueAccent] (fb_up) to[out=90, in=-90] (bubble_feedback.south);
\draw[accent, color=blueAccent] (perf_left) -- (fb_right);
\draw[accent, color=black!40] (cand_down) -- (perf_up);

\path (fb_up) -- (bubble_feedback.south)
  node[midway, right, text=blueAccent, font=\bfseries, xshift=0.2cm] {Feedback};
\path (perf_left) -- (fb_right)
  node[midway, below, text=blueAccent, font=\bfseries, yshift=-0.05cm] {Performance};

\end{tikzpicture}
\caption{Agentic two-loop workflow for SystemC model generation: Loop~A synthesizes candidate model code from task and context, and Loop~B refines it using deterministic feedback.}
\label{fig:workflow}
\end{figure*}

For full ECUs, the feedback mechanism is applied hierarchically: components first, then subsystems, and finally the integrated system. This staged calibration supports convergence toward consistent, reproducible, and hardware-near virtual ECU twins. Figure~\ref{fig:loops} summarizes the two-loop calibration cycle in a compact schematic, while Figure~\ref{fig:workflow} illustrates a concrete instantiation used for CPU-model generation.

\section{Case Study: Automated CPU Modeling}
The prototype focused on the most critical ECU component: the processor. A code agent generated a SystemC/TLM~2.0 CPU model and iteratively improved it via differential testing against a trusted ARMv8 reference instruction-set simulator accessed through a GNU Debugger (GDB)-compatible interface. The exact processor instance is not disclosed, because the validation target of this work is ISA-level behavioral correctness rather than processor-specific microarchitectural detail.

\subsection{Instruction-Level Validation Strategy}
Validation was performed at the architecturally visible instruction-set level. For each instruction, the expected behavior was defined by the ARMv8 ISA specification, while the reference simulator provided an executable, deterministic baseline for automated comparison. The simulator therefore acted as a practical ground truth for regression and discrepancy detection, whereas the ISA documentation defined the intended semantics.

To test individual instructions, short self-contained test programs were constructed that isolated the behavior of a single target instruction under controlled preconditions. For example, a \texttt{MOV} instruction was exercised by initializing source and destination registers to known values, executing the instruction in both the reference simulator and the generated SystemC model, and then comparing the resulting architectural state. The comparison covered the relevant general-purpose registers, condition flags where applicable, program-counter progression, and any instruction-visible memory effects.

This procedure was repeated across the full ARMv8 ISA at the architectural level. The methodological novelty is therefore not a model of one specific processor instance, but a generative feedback loop that synthesizes candidate behavior, validates it against reference execution, and repairs mismatches until the generated model converges toward ISA-consistent behavior. Operationally, the refinement loop was:
\begin{enumerate}
  \item Execute an instruction in the reference model.
  \item Read architectural state via GDB (registers, flags).
  \item Execute the same instruction in the generated SystemC model.
  \item Compare states; feed discrepancies back to the agent for correction.
\end{enumerate}

This closed-loop process enables instruction-accurate behavior without manually implementing every corner case upfront. The project demonstrated that an agent can converge on instruction set architecture (ISA)-conformant behavior across the full ARMv8 ISA at the architectural level.

\section{Results}
The project produced a working prototype that validated instruction-accurate CPU modeling and automated correction. Key qualitative outcomes were:
\begin{itemize}
  \item Instruction-accurate CPU execution across the full ARMv8 ISA at the architectural level with full architectural-state observability.
  \item Automated discrepancy detection and iterative correction based on register/flag deltas.
  \item Compatibility with non-intrusive tracing and fault-injection campaigns relevant to ISO~26262-style robustness analysis.
  \item Reduced dependency on scarce hardware-in-the-loop resources by enabling repeatable host-based execution and analysis.
\end{itemize}

Prototypical, lab-scale implementations further indicate that automatically generated model structures can be calibrated iteratively toward observed hardware behavior. In particular, experiments showed that:
\begin{itemize}
  \item Deterministic register and communication sequences can be reproduced in the virtual environment.
  \item Trace deviations between real and simulated execution can be reduced over multiple iterations.
  \item Individual peripheral components can be integrated consistently into a virtual system model.
\end{itemize}

These outcomes support the core claim: a feasible path exists to build ECU twins capable of running real binaries early in the development cycle, enabling integration and safety testing before physical prototypes are available.

\section{Discussion and Limitations}
The project intentionally prioritized technical feasibility over platform-scale deployment. As a result, cloud multi-tenancy, CI/CD pipelines, and large-scale regression automation remain out of scope. Another limitation is peripheral fidelity: while CPU correctness was addressed, the breadth of ECU-specific peripherals and real-time constraints must be modeled to ensure behavioral equivalence at the system level. Finally, safety evidence generation and audit trails remain emerging areas, even though fault injection is conceptually supported.

Conceptually, the contribution is a closed-loop workflow that couples (i) generative synthesis of model code with (ii) deterministic differential evaluation against a reference and (iii) structured feedback that drives iterative repair. Unlike manual expert calibration or workflows dependent on proprietary vendor data, the method uses data-driven iteration to improve model fidelity. In contrast to EUREKA, which optimizes reward code for learned policies, the optimization target here is deterministic simulation behavior under physical, timing, and safety constraints. The resulting evaluation space integrates timing, resource usage, interference behavior, fault responses, and determinism---providing a foundation for audit-ready safety artifacts aligned with ISO~26262-style evidence expectations.

From a legal and ethical standpoint, the workflow presupposes authorized access to target hardware and does not rely on bypassing security controls or redistributing proprietary artifacts; practitioners should apply it in compliance with applicable intellectual-property, contractual, and anti-circumvention requirements.

An additional limitation concerns simulation scalability in standard SystemC environments. The Accellera reference implementation is based on a discrete-event simulation kernel with a cooperative process scheduler and typically executes within a single host thread~\cite{systemc}. As a result, even if a model represents a 64-core SoC, the simulation often runs primarily on one CPU core of the host machine. This can become a bottleneck for time-accurate vECUs and large-scale regression workloads.

\section{Conclusion and Outlook}
The InnoRegioChallenge 2025 project demonstrates that instruction-accurate CPU models can be generated and refined automatically using a reference simulator and GDB-based differential testing. This reduces a key technical risk in virtual ECU development and establishes a foundation for shift-left testing, fault injection, and scalable regression. While platform industrialization is still required, the results show a credible path toward practical, high-fidelity vECUs for automotive software development.

Future work will extend the agentic synthesis and feedback-calibration workflow beyond CPUs to additional ECU components. This includes GPUs and dedicated hardware accelerators, standardized integrated circuits (e.g., power management, timers, transceivers), and analog subsystems (e.g., filtering and sensor front-ends) via SystemC-AMS or SPICE-compatible models. A key goal is to scale the approach hierarchically from component-level calibration to subsystem and full-ECU convergence while preserving determinism, timing fidelity, and safety-relevant observability.

To mitigate SystemC performance limitations in larger vECU setups, several engineering paths are promising:
\begin{itemize}
  \item \textbf{Parallel SystemC implementations:} research and commercial tools provide parallel discrete-event simulation (PDES), multi-threaded kernels, and partitioned simulation; these capabilities are not part of the standard open-source reference implementation.
  \item \textbf{Multi-process co-simulation:} partition the platform and run components in separate processes, communicating via IPC or transaction-level interfaces; this can utilize multiple host CPU cores while keeping model boundaries explicit.
  \item \textbf{Hybrid C++ threading:} leverage native threads (e.g., \texttt{std::thread} or OpenMP) for computations around the model, while carefully synchronizing with the SystemC kernel; the kernel itself remains single-threaded.
\end{itemize}

\section*{Acknowledgment}
This work was supported by a grant from Wirtschaftsf\"orderung Region Stuttgart GmbH (WRS) for conducting InnoRegioChallenge 2025.

\section*{Transparency Statement}
Transparency on generative AI use: For manuscript preparation, we employed ChatGPT (OpenAI, GPT-5.2, accessed February 2026) to support content development (brainstorming alternatives, generating outlines, and rephrasing text for clarity). In this manuscript-preparation context, the tool was not used to conduct analyses, create results, or verify factual claims. Separately, the use of Claude Opus~4-class coding models as part of the research workflow for SystemC/TLM code generation and repair is disclosed in the Method section. All factual statements, equations, figures, tables, and citations were produced and validated by the authors, who accept full responsibility for the content.

\end{document}